\documentclass[pra,amsmath,showpacs%
,twocolumn%
,superscriptaddress,letterpaper]{revtex4}
\usepackage{graphicx}
\newcommand{\bra}[1]{\left\langle#1\right|}
\newcommand{\ket}[1]{\left|#1\right\rangle}

\newcommand{\scp}[2]{\langle #1\!\mid\!#2\rangle}
\newcommand{\ave}[1]{\langle#1\rangle}
\newcommand{\fs}{{\tilde f}}
\newcommand{\ofs}{(1+\tilde f)}
\newcommand{\ms}{{\tilde m}}
\newcommand{\Gs}{\tilde G}

\newcommand{\ofsp}{(1+\tilde f)^{\vphantom\ast}}
\newcommand{\qandq}{\quad\text{and}\quad}
\newcommand{\hc}{\mathrm{H.c.}}

\newcommand{\der}[1]{\frac{d}{d#1}}
\newcommand{\amp}{\!\!&\!\!}
\newcommand{\Hzero}{\hat{H}^{(0)}}
\newcommand{\ie}{i.\thinspace e., }
\newcommand{\ant}{~and~}
\newcommand{\Sig}[1]{\Sigma_{\mathcal{#1}}}
\newcommand{\Pim}[1]{\Pi_{\mathcal{#1}}}
\newcommand{\mc}{m^{c}}
\newcommand{\fc}{f^{c}}
\newcommand{\Eno}{\epsilon^{0}}
\newcommand{\Esp}{\epsilon^{\text{sp}}}
\newcommand{\ea}{~{\it et al.}}
\newcommand{\opp}{(\sigma_3+P)}
\newcommand{\Gam}[1]{\Gamma_{\mathcal{#1}}}
\newcommand{\LandO}{\mathcal{O}}
\newcommand{\De}{\Delta\epsilon}
\newcommand{\otfsp}{(1+2\tilde f)^{\vphantom\ast}}
\newcommand{\T}{T^{\vphantom<}}
\newcommand{\tr}{{\mathrm{Tr}}}
\newcommand{\bx}{{\mathbf{x}}^{\vphantom\dag}}
\newcommand{\bp}{{\mathbf{p}}}
\newcommand{\Pv}{P^{\vphantom\ast}}
\newcommand{\fsp}{\tilde f^{\vphantom\ast}}
\newcommand{\m}{m^{\vphantom\ast}}
\newcommand{\V}[1]{V_{\mathrm{#1}}}
\newcommand{\al}{\alpha^{\vphantom\ast}}
\newcommand{\vp}{{\vphantom{\tilde f}}}
\newcommand{\f}{f^{\vphantom\ast}}
\newcommand{\av}{\hat{a}^{\vphantom\dag}}
\newcommand{\adag}{\hat{a}^\dag}
\newcommand{\msp}{\tilde m^{\vphantom\ast}}

\begin{document}
  \title{Gapless kinetic theory beyond the Popov approximation}
  \author{J.~Wachter}
  \affiliation{JILA, National Institute of Standards and Technology and
	University of Colorado, \\Boulder, Colorado 80309-0440, USA}
  \author{R.~Walser}
  \affiliation{Universit\"at Ulm, Abteilung f\"ur Quantenphysik, 
	D-89069 Ulm, Germany}
  \author{J.~Cooper}
  \author{M.~Holland}
  \affiliation{JILA, National Institute of Standards and Technology and
	University of Colorado, \\Boulder, Colorado 80309-0440, USA}
  \date{December 17, 2002, submitted to PRA}
  \begin{abstract}

We present a unified kinetic theory that describes the
finite-temperature, non-equilibrium dynamics of a Bose-Einstein
condensed gas interacting with a thermal cloud. This theory includes
binary interactions to second order in the interaction potential and
reduces to a diagonal quantum Boltzmann equation for Bogoliubov
quasiparticles.  The Hartree-Fock-Bogoliubov interactions include the
pairing field and are expressed as many-body $T$ matrices to second
order. The interactions thus include the correct renormalized
scattering physics.  This renormalized theory is automatically
gapless.  Thus, the excited Bogoliubov modes are naturally orthogonal
to the condensate ground state.  \end{abstract}



\pacs{03.75.Kk, 05.70.Ln} 
\keywords{Bose-Einstein condensation, kinetic theory, renormalization}
  \maketitle

\section{Introduction}

Finite-temperature theories of Bose-Einstein condensation (BEC) have
been a very active field of study. The goal is a unified description
of a dilute, atomic gas of bosons in a harmonic trap in terms of a
condensate mean field interacting with a thermal cloud.  The success
of the zero-temperature Gross-Pitaevskii (GP) theory in describing BEC
experiments spurred interest in effects not contained in this
framework. Examples are damping of collective excitations and
condensate growth through collisional redistributions of thermal
atoms. This redistribution cannot be treated in theories that are
still of first order in the binary interaction, such as, for example,
Hartree-Fock-Bogoliubov (HFB) theories.

We present a kinetic theory of second order in the interaction, which
is formulated in terms of Bogoliubov quasiparticles and contains
collisional terms beyond the HFB approximation. The HFB interactions
are expressed as many-body $T$ matrices to second order in the binary
potential and thus include the correct renormalized scattering
physics.  This theory thus contains no ultraviolet divergences and has
a gapless energy spectrum. We extend the papers of
Walser\ea~\cite{Walser1999a,Walser2001a}, which are the basis of the
present results, by these two essential aspects.

We go beyond theoretical approaches that drop the anomalous pair matrix
~$\ms=\ave{\tilde a\tilde a}$ in the Popov approximation
\cite{Zaremba1999a,AlKhawaja2000a,Imamovic2000b,Graham2000b}. Monte
Carlo simulations based on the semi-classical Zaremba-Nikuni-Griffin
theory~\cite{Zaremba1999a} show very good agreement for experimentally
observed damping rates and response
frequencies~\cite{Jackson2001a,Jackson2002c}. However, recent
experiments~\cite{Donley2002a} and their theoretical
explanations~\cite{Kokkelmans2002b,Koehler2002b} have shown that the
pairing field plays an important role in Bose gases with resonance
interactions.

Keeping the pair matrix~$\ms$ in this theory would cause ultraviolet
divergences if we replaced the bare interaction potentials with the
$s$-wave scattering length $a_s$ in the contact-potential
approximation. The vacuum part of the pairing field's self
interaction, for example, will diverge, because the delta-function
potential contains unphysically high energy contributions. These
divergences can be resolved by writing the interactions in terms of
scattering~$T$ matrices, which subsume the divergent sums over
intermediate scattering states and correctly reduce to the scattering
length~$a_s$ in the zero-energy and -momentum
limit~\cite{Proukakis1998a,Stoof1999a,Hutchinson2000a,Morgan2000a,Koehler2002a}.
Using these scattering~$T$ matrices, we can consistently eliminate all
divergences to second order in the interaction.

We thus obtain a renormalized HFB operator, which has a gapless
spectrum. The zero-energy eigenspace is spanned by the condensate, and
excitations with non-vanishing energy are thus automatically
orthogonal to the condensate. This is similar to the renormalized
gapless HFB equations proposed in Ref.~\cite{Hutchinson2000a}. Other
approaches have to explicitly project the excitations orthogonal to
the condensate \cite{Morgan2000a,Castin1997a}. Using the condensate as
the ground state of this adiabatic basis also simplifies the
complicated representation of a condensate band in the Gardiner-Zoller
master equation formulation \cite{Davis2000a}.

The Monte Carlo simulations of Jackson and Zaremba
\cite{Jackson2001a,Jackson2002c} show that considering dynamic
population-ex\-change between the condensate and the thermal cloud and
within the thermal cloud leads to good agreement with the observed
response spectra. We include these kinetic effects, thus going beyond
collisionless descriptions \cite{Giorgini2000a,Fedichev1998b}.

Our presentation builds on the papers by Walser {\it et al.}\
\cite{Walser1999a,Walser2001a} and we will begin by summarizing the
kinetic equations derived in these papers in Sec.~\ref{spke}. In
Sec.~\ref{ren}, we examine the first-order HFB propagator and rewrite
it in terms of second-order $T$ matrices, by including second-order
energy shifts and adiabatically eliminating the pairing
field~$\ms$. This shows that the theory is explicitly gapless and
renormalized. We then make use of the gaplessness and write the
kinetic equations in terms of Bogoliubov quasiparticles, which are
orthogonal to the condensate by construction. The results in
Sec.~\ref{qke} show that the complicated collision terms presented in
the Walser {\it et al.}\ papers can be simplified dramatically by a
basis transformation. Practical calculations of the quantum Boltzmann
equation then will require only diagonal elements of the quasiparticle
population matrix.

\section{Single-Particle Kinetic Equations}\label{spke}

We present the Walser {\it et al.}\ formulation of kinetic theory
\cite{Walser2001a}, which was originally derived using a statistical
operator approach \cite{Walser1999a}. The authors of this paper
also recently showed a derivation of the same equations
\cite{Wachter2001a} from the Kadanoff-Baym \cite{Kadanoff,Kane1965a}
theory of non-equilibrium Green functions. They used the gapless
second-order Beliaev approximation \cite{Imamovic2000b,Beliaev1958b}
and showed equivalence to the work of Walser {\it et al.}\ under the
Markov approximation. Recently, another independent derivation
\cite{Proukakis2001a,Proukakis2002a} connected this approach to
kinetic theories by Morgan and Proukakis.

Neglecting three-body and higher interactions, we can describe the
weakly interacting, dilute gas by the following Hamiltonian
\begin{equation}
\hat{H}={H^{(0)}}^{1'2'}\adag_{1'}\av_{2'} + 
        \phi^{1'2'3'4'}\adag_{1'}\adag_{2'}\av_{3'}\av_{4'},
\end{equation}
where ${H^{(0)}}^{1'2'}=\bra{1'}\Hzero\ket{2'}$ denotes the matrix
elements of the interaction-free single-particle Hamiltonian
\begin{equation}
  \Hzero=\frac{\hat{\bp}^2}{2m}+\V{ext}(\hat{\bx})	\label{def_Hzero}
\end{equation}
with external harmonic potential~$\V{ext}$. The bosonic creation
operator~$\adag_{1'}$ creates a particle in the state~$\ket{1'}$,
where $1'$ stands for a complete set of quantum numbers, which label a
constant, single-particle energy basis, such as, for example, harmonic
oscillator states or eigenstates of the GP equation. We use the
summation convention for these abbreviated indices and indicate the
single-particle basis with primes.

The two-particle matrix elements of the binary interaction
potential~$\V{bin}(\bx_1,\bx_2)$ are defined by 
\begin{eqnarray}
\phi^{1'2'3'4'}=\frac14\amp\displaystyle\int\amp d\bx_1\; d\bx_2\; 
        \scp{1'}{\bx_1}\scp{2'}{\bx_2}\V{bin}(\bx_1,\bx_2)\nonumber\\
        \amp\times\amp\Big\{\scp{\bx_1}{3'}\scp{\bx_2}{4'}+
	\scp{\bx_1}{4'}\scp{\bx_2}{3'}\Big\}\label{defphi}.
\end{eqnarray}
These matrix elements are symmetric in the first and last two indices:
\begin{equation}
  \phi^{1'2'3'4'}=\phi^{2'1'3'4'}=\phi^{1'2'4'3'}.	\label{phisymm}
\end{equation}

To determine the measurable quantities we want to calculate in this
theory, we first define the condensate mean field~$\alpha$ as the
expectation value of the destruction operator
\begin{equation}
  \alpha=\al_{1'}\ket{1'}\equiv\ave{\av_{1'}}\ket{1'}.	\label{defal}
\end{equation}
The total density matrix~$f$ is defined by
\begin{equation}
  f\equiv\ave{\adag_{2'}\av_{1'}}\ket{1'}\otimes\bra{2'}.
\end{equation}
Subtracting the condensate density matrix
\begin{equation}
  f^c\equiv\alpha\otimes\alpha^\ast
	\equiv\alpha^\ast_{2'}\al_{1'}\ket{1'}\otimes\bra{2'},
\end{equation}
we obtain the density matrix of thermal atoms~$\fs\equiv f-f^c$. The
anomalous average~$m$ is split analogously
\begin{equation}
  m=\ave{\av_{1'}\av_{2'}}\ket{1'}\otimes\ket{2'}=m^c+\ms,
\end{equation}
in the condensate part~$m^c\equiv\alpha\otimes\alpha$ and
fluctuations~$\ms$.

This set of variables contains all possible combinations of up to two
field operators. The course-grained statistical operator parametrized
by these variables is thus Gaussian, and we can use Wick's theorem to
truncate the coupling to higher-order correlation functions,
\ie expectation values of more then two field operators.
This approximation is valid because of the diluteness of the condensed
gas. In a dilute gas the duration of a collision event is short
compared to the essentially free evolution between collisions, which
allows higher-order correlations to damp.

The procedure followed by Walser {\it et al.}\ \cite{Walser1999a} is
then to write the Heisenberg equations of motion for the variables
above and expand the expectation values using Wick's theorem. Walser
{\it et al.}\ thus obtain the equations of motion given in the
following two sections \cite{Walser2001a}.
\subsection{Mean-field equations}
Since the anomalous fluctuations~$\ms$ couple the mean field~$\alpha$
to its conjugate~$\alpha^\ast=\alpha^\ast_{1'}\bra{1'}$, it is
convenient to write the generalized Gross-Pitaevskii (GP) equation in
a two-by-two matrix form 
\begin{equation}
  \der{t}\chi=(-i\Pi+\Upsilon^<-\Upsilon^>)\chi,	\label{ggpe}
\end{equation}
where the two-component state vector
\begin{equation}
\chi=\left(\begin{array}{c}\alpha\\\alpha^\ast\end{array}\right)
	\label{defchi}
\end{equation}
is defined in terms of $\alpha=\al_{1'}\ket{1'}$ and also contains the
time-reversed mean field~$\alpha^\ast$.  

The generalized GP propagator representing the reversible evolution
of~$\chi$ is defined as
\begin{equation}
\Pi=\left(\begin{array}{cc}
	\Pi_{\cal N}&\Pi_{\cal A}\\
	-\Pi_{\cal A}^\ast&-\Pi_{\cal N}^\ast\end{array}\right).
	\label{defpi}
\end{equation}
This symplectic propagator consists of the normal Hermitian Hamiltonian 
\begin{equation}
  \Pi_{\cal N}=\Hzero+U_{f^c}+2U_{\fs}-\mu,	\label{defpin}
\end{equation}
where~$\mu$ removes rapid oscillations of the mean field,
and the symmetric anomalous coupling
\begin{equation}
  \Pi_{\cal A}=V_{\ms}.	\label{defpia}
\end{equation}
The energy shifts due to both the mean field and the normal fluctuations are
given by the matrices
\begin{equation}
  U_{f}=2\;\phi^{1'2'3'4'}\f_{3'2'}\ket{1'}\otimes\bra{4'},\label{defUf}
\end{equation}
whereas the first-order anomalous coupling-strength is given by
\begin{equation}
  V_{m}=2\;\phi^{1'2'3'4'}\m_{3'4'}\ket{1'}\otimes\ket{2'}. \label{defVm}
\end{equation}

\begin{figure}[t]
  \centering{\includegraphics[scale=.8]{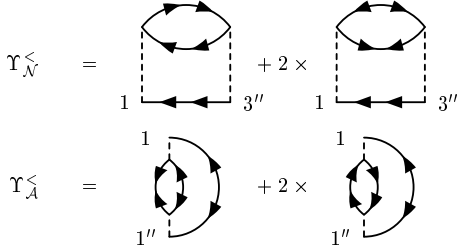}}
  \caption{The diagrams corresponding to the second-order
  terms~$\Upsilon^<$ in the GP~equation \eqref{ggpe}. The dashed
  potential lines correspond to the symmetrized binary
  potential~$\phi$ in the single-particle energy basis. The
  directed propagators represent the normal density~$\fs$, the
  remaining ones the anomalous average~$\ms$ and its conjugate. 
  Note that all diagrams are topologically equivalent, and only
  propagators are exchanged.
  \label{fig_gamma}}
\end{figure}

The second-order irreversible evolution, consisting of damping rates
and energy shifts, is given by the collision operator
\begin{equation}
\Upsilon^<=\begin{pmatrix}\Upsilon^<_{\mathcal{N}}&\Upsilon^<_{\mathcal{A}}\\
	-\Upsilon^{>\ast}_{\mathcal{A}}&-\Upsilon^{>\ast}_{\mathcal{N}}
	   \end{pmatrix}	\label{defup}
\end{equation}
and its time-reversed
counterpart~$\Upsilon^>=-\sigma_1\Upsilon^{<\ast}\sigma_1$,
where~$\sigma_1$ is the first Pauli matrix exchanging the positive-
and negative-energy components of vectors. The matrix elements of the
collision operator are given by
\begin{subequations}\label{defups}
\begin{eqnarray}
  \Upsilon^<_{\mathcal{N}}&=&\Gamma_{\fs\fs\ofs}+2\Gamma_{\fs\ms\ms^\ast},\\
  \Upsilon^>_{\mathcal{N}}&=&\Gamma_{\ofs\ofs\fs}+2\Gamma_{\ofs\ms\ms^\ast},\\
  \Upsilon^<_{\mathcal{A}}&=&\Gamma_{\vp\ms\ms\ms^\ast}+2\Gamma_{\fs\ms\ofs},\\
  \Upsilon^>_{\mathcal{A}}&=&\Gamma_{\vp\ms\ms\ms^\ast}+2\Gamma_{\ofs\ms\fs},
\end{eqnarray}
\end{subequations}
which is defined in terms of individual collisions~$\Gamma$.
These elementary collision processes are defined explicitly as
\begin{widetext}
\begin{subequations}\label{defgamma}
\begin{eqnarray}
  \Gamma_{fff}&=&8\phi^{1'2'3'4'}\phi_{\eta}^{1''2''3''4''}
	\f_{3'1''}\f_{4'2''}\f_{4''2'}\ket{1'}\otimes\bra{3''},\\
  \Gamma_{fmf}&=&8\phi^{1'2'3'4'}\phi_{\eta}^{1''2''3''4''}
	\f_{3'2''}\m_{4'3''}\f_{4''2'}\ket{1'}\otimes\ket{1''},\\
  \Gamma_{fmm^\ast}&=&8\phi^{1'2'3'4'}\phi_{\eta}^{1''2''3''4''}
	\f_{3'1''}\m_{4'4''}m^\ast_{2''2'}\ket{1'}\otimes\bra{3''},\\
  \Gamma_{mmm^\ast}&=&8\phi^{1'2'3'4'}\phi_{\eta}^{1''2''3''4''}
	\m_{3'4''}\m_{4'3''}m^\ast_{2''2'}\ket{1'}\otimes\ket{1''}.
\end{eqnarray}
\end{subequations}
\end{widetext}
The in-rates of the collision operator~$\Upsilon$ are depicted in
Fig.~\ref{fig_gamma}.

In this theory, collisional interactions are considered to second
order. The effect of higher order terms, which lead to a finite
duration of a collision, can be modeled by introducing a
parameter~$\eta$, such that every second-order collision operator
contains dispersive as well as dissipative parts from the
complex-valued matrix element
\begin{equation}
  \phi_{\eta}^{1''2''3''4''}= \phi^{1''2''3''4''}\frac1{\eta-i\De},
\end{equation}
where the energy
difference~$\De=-(\Eno_{1''}+\Eno_{2''})+\Eno_{3''}+\Eno_{4''}$ has to
be smaller than the energy uncertainty~$\eta$ to get a sizable
contribution. Note that the papers \cite{Walser1999a,Walser2001a}
contain a sign error in the definition of~$\De$. For small $\eta$ we
obtain
\begin{equation}
  \frac{1}{\eta-i\De}
	\quad\underset{\eta\rightarrow0}{\longrightarrow}\quad
	\pi\delta_\eta(\De)+i{\mathcal P}_\eta
	\frac{1}{\De}, \label{principalspe}
\end{equation}
where $\mathcal P$ indicates that the Cauchy principal value has to be
taken upon integration.  The parameter~$\eta$ thus represents
off-the-energy-shell propagation after a collision. Most
off-the-energy-shell coherences decay during subsequent propagation,
but, due to the finite time between collisions~$\tau_c$, energy cannot
be conserved exactly, because $\eta$ has to be larger than the
collision rate~$1/\tau_c$. 

\subsection{Equations for normal densities and anomalous fluctuations
		\label{jkt_fl}}
The equations of motion for the fluctuation densities $\fs$~and~$\ms$
are coupled and can also conveniently be written in terms of
two-by-two matrices. To achieve this, we define the generalized
single-time fluctuation-density matrix~$\Gs^<$ as
\begin{equation}
  \Gs^<=\left(\begin{array}{cc}\fs&\ms\\
	\ms^\ast&\ofs^\ast\end{array}\right),  \label{def_gsl}
\end{equation}
where
$\fs=\fs_{1'2'}\ket{1'}\otimes\bra{2'}$~and~$\ms=\ms_{1'2'}\ket{1'}\otimes\ket{2'}$
are the matrix representations of the master variables.  In
Ref.~\cite{Wachter2001a}, we use the property that this density matrix
is the single-time limit of the corresponding time-ordered two-time
Green's function. This showed that the other time ordering is
given by
\begin{equation}
  \Gs^>=\left(\begin{array}{cc}\ofs&\ms\\\ms^\ast&\fs^\ast\end{array}\right)
	=\sigma_1\Gs^{<\ast}\sigma_1=\sigma_3+\Gs^<.	\label{def_gsg}
\end{equation}
Here, we use the third Pauli
matrix~$\sigma_3=\mathrm{diag}(\openone,-\openone)$. Note that our
naming of the fluctuation-density matrices $\Gs^<$\ant$\Gs^>$ is
consistent with the two-time formalism in Ref.\ \cite{Wachter2001a},
but disagrees with Ref.\ \cite{Walser2001a}.

The generalized Boltzmann equation of motion for this
fluctuation-density matrix can be written as
\begin{equation}
  \der{t}\Gs^<=-i\Sigma\Gs^<+\Gamma^<\Gs^>-\Gamma^>\Gs^<+\hc
	\label{gbe}
\end{equation}
This equation has to be solved under the constraints
\begin{equation}
  \alpha^\ast\fs=0\qandq\alpha^\ast\ms=0,
	\label{alphaGort}
\end{equation}
which force the fluctuations to be orthogonal to the condensate.

Again, the equation of motion~\eqref{gbe} has two parts: The
reversible evolution is governed by the Hartree-Fock-Bogoliubov
self-energy operator
\begin{equation}
  \Sigma=\left(\begin{array}{cc}\Sig{N}&\Sig{A}\\
	-\Sig{A}^\ast&-\Sig{N}^\ast\end{array}\right),	\label{defsig}
\end{equation}
which in turn consists of the Hermitian Hamiltonian
\begin{equation}
  \Sig{N}=\Hzero+2U_{f^c}+2U_{\fs}-\mu	\label{defsign}
\end{equation}
and the symmetric anomalous coupling
\begin{equation}
  \Sigma_{\cal A}=V_{\mc}+V_{\ms}.	\label{defsiga}
\end{equation}

The irreversible evolution introduced by second-order collisional
contributions now consists of the collisional operator
\begin{equation}
  \Gamma^<=\begin{pmatrix}\Gam N^<&\Gam A^<\\
	-\Gam A^{>\ast}&-\Gam N^{>\ast}\end{pmatrix},  \label{defgam}
\end{equation}
and its time-reversed counter part
$\Gamma^>=-\sigma_1\Gamma^{<\ast}\sigma_1$. The diagonal components of
the collision operator are defined as 
\begin{eqnarray}
  \Gam N^<&=&\Gamma_{(\fs+\fc)\fs\ofs}+\Gamma_{\fs\fc\ofs}
	+\Gamma_{\fs\fs\fc}\label{defgamn}\\
  &&+2\big\{\Gamma_{(\fs+\fc)\ms\ms^\ast}+\Gamma_{\fs\mc\ms^\ast}
      +\Gamma_{\fs\ms m^{c\ast}}\big\},\nonumber\\
  \Gam N^>&=&\Gamma_{(1+\fs+\fc)\ofs\fs}+\Gamma_{\ofs\fc\fs}
      +\Gamma_{\ofs\ofs\fc}\\
  &&+2\big\{\Gamma_{(1+\fs+\fc)\ms\ms^{\ast}}+\Gamma_{\ofs\mc\ms^\ast}
	+\Gamma_{\ofs\ms m^{c\ast}}\big\}\nonumber,
\end{eqnarray}
and the off-diagonal, anomalous components as
\begin{eqnarray}
  \Gam A^<&=&\Gamma_{\vp(\ms+\mc)\ms\ms^\ast}+\Gamma_{\vp\ms\mc\ms^\ast}
      +\Gamma_{\vp\ms\ms m^{c\ast}}\label{defgama}\\
  &&+2\big\{\Gamma_{(\fs+\fc)\ms\ofs}+\Gamma_{\fs\mc\ofs}
      +\Gamma_{\fs\ms\fc}\big\},\nonumber\\
  \Gam A^>&=&\Gamma_{\vp(\ms+\mc)\ms\ms^\ast}+\Gamma_{\vp\ms\mc\ms^\ast}
      +\Gamma_{\vp\ms\ms m^{c\ast}}\\
  &&+2\big\{\Gamma_{(1+\fs+\fc)\ms\fs}+\Gamma_{\ofs\mc\fs}
      +\Gamma_{\ofs\ms\fc}\big\}.\nonumber
\end{eqnarray}

\begin{widetext}
\
\begin{figure}[h]
  \parbox{6cm}{\caption{The diagrams corresponding to the second-order
  terms~$\Gamma^<$ in the generalized Boltzmann equation
  \eqref{gbe}. The dashed lines depict the symmetrized binary
  potential~$\phi$ in the single-particle energy basis. The directed
  propagators represent the normal density~$\fs$, the remaining ones
  the anomalous average~$\ms$ and its conjugate. The first column of
  diagrams is identical to those depicted in Fig.~\ref{fig_gamma}. The
  remaining diagrams each have one of the three propagators replaced
  with an open condensate line.  \label{fig_gammasq}}} \hfill
  \parbox{10.3cm}{\includegraphics[scale=.8]{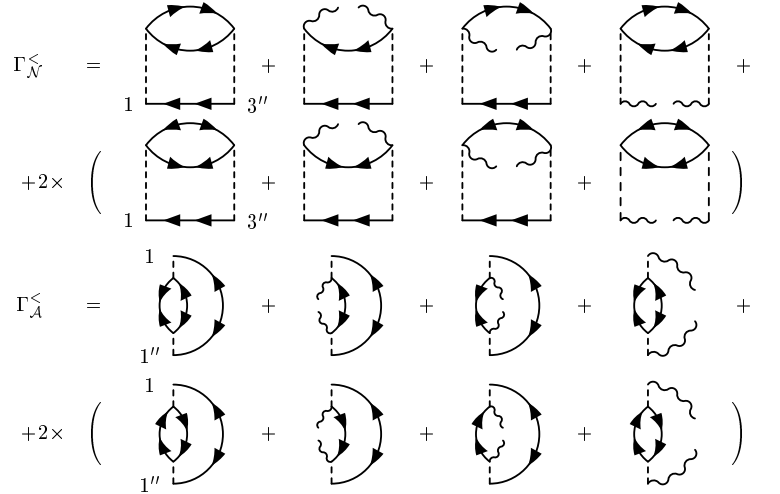}}
\end{figure}
\end{widetext}

Both the diagonal and off-diagonal incoming rates are depicted
diagrammatically in Fig.~\ref{fig_gammasq}.
For every term that appears in the collisional terms of the
generalized GP equation in~$\Upsilon^<$ [Fig.~\eqref{fig_gamma}], we
here [Fig.~\eqref{fig_gammasq}] have three additional terms, where in
each of them, one of the three fluctuating contributions is replaced
with the corresponding mean-field quantity. This replacement rule can
be seen in the Beliaev collisional self energies presented in
Ref.~\cite{Wachter2001a} and is a consequence of the fact that the
Boltzmann equation~\eqref{gbe} can be generated from the GP
equation~\eqref{ggpe} by functional differentiation
\cite{Hohenberg1965a}.

When the collision operator~$\Gamma^<$ is multiplied by~$\Gs^>$ as in
Eq.~\eqref{gbe}, we get terms like
\begin{equation}
  \Gamma_{\fs\fs\ofs}\ofs-\Gamma_{\ofs\ofs\fs}\fs,
\end{equation}
where the second part comes from the time-reversed
contribution~$\Gamma^>\Gs^<$. The diagonal parts are exactly the in
and out terms of the quantum Boltzmann equation for the
single-particle distribution function~$\fs$. The remaining second-order
contributions couple to the anomalous fluctuations~$\ms$ and do not
have an analogue in the quantum Boltzmann equation. In Sec.~\ref{qke},
we will rewrite the kinetic equations presented here in terms of
Bogoliubov quasiparticles. Then all collisional contribution take the
form of Boltzmann terms. 

\section{Gaplessness --- $T$ matrices}\label{ren}

Our goal in this section is to explicitly show that the kinetic
equations \eqref{gbe}\ant\eqref{ggpe} are gapless. This should on one
hand be obvious, because our previous paper~\cite{Wachter2001a} showed
them to be equivalent to the Kadanoff-Baym equations~\cite{Kadanoff,
Kane1965a} in the gapless Beliaev
approximation~\cite{Imamovic2000b,Beliaev1958b}. The first-order HFB
propagator~$\Sigma$, which appears in the kinetic equations, is, on
the other hand, known to exhibit a non-physical energy gap in the
long-wavelength, homogeneous limit \cite{Griffin1996a}.

We resolve this discrepancy by including second-order collisional
energy shifts $\mathcal{P}\{\Gamma\}$ into the HFB propagator and
adiabatically eliminating the anomalous average~$\ms$ in the
first-order anomalous potential~$V_{\ms}$ \eqref{defVm}. This upgrades
the bare interaction potentials in the first-order propagators
$\Sigma$\ant$\Pi$ to the real parts of many-body $T$ matrices. We can
then approximate these $T$ matrices with the $s$-wave scattering length
without incurring ultra-violet divergences.

The upgraded HFB propagator~$\Sigma$, where all binary interactions
are written as many-body $T$ matrices, is explicitly gapless and thus
obeys the Hugenholtz-Pines theorem \cite{Hugenholtz1959a}. The
propagator thus has zero-energy modes, which are completely specified
by the value of the condensate~$\alpha$. If we use the non-zero energy
Bogoliubov modes of $\Sigma$ as a basis for the thermal excitations,
the excitations will automatically be orthogonal to the condensate. We
follow this idea in Sec.~\ref{qke}.

\subsection{Off-diagonal potentials}
Here, we update the off-diagonal potentials~$V_{(\mc+\ms)}$ in the HFB
propagator~$\Sigma$ by adiabatically eliminating the pairing
field~$\ms$.  We integrate the first-order equation of motion for the
anomalous average~$\ms$
\begin{equation}
  \der t\ms=-i\Sig N\ms-i\ms\Sig N-i\Sig A\ofs^\ast-i\fs\Sig A,
\end{equation}
which is obtained by taking the $\ms$ component of the generalized
Boltzmann equation \eqref{gbe} and dropping the second-order terms,
because we want to substitute the result for $\ms$ into the anomalous
potential~$V_\ms$ and only keep terms up to second order. 
In stationarity, we solve for~$\ms$ in the eigenbasis of~$\Sig{N}$,
\begin{equation}
  \Sig{N}\ket{\Eno_{1'}}=(\Eno_{1'}-\mu)\ket{\Eno_{1'}},	\label{Emu}
\end{equation}
and obtain
\begin{equation}
  \msp_{1'2'}=\mathcal{P}\frac{\Sig A^{1'2''}\ofsp_{2'2''}+
	\fsp_{1'2''}\Sig A^{2''2'}}{2\mu-(\Eno_{1'}+\Eno_{2'})}	\label{mphi}
\end{equation}
as an adiabatic solution. Adiabatic here means that this solution only
includes time-variations with characteristic times long compared to
the duration of a collision. We use the Cauchy principal
value~$\mathcal{P}$ to indicate omission of the divergent term in an
energy integral or sum. This divergent $\delta$-function term gives
rise to the imaginary part. We plug this result into the off-diagonal
potential \eqref{defVm},
\begin{equation}
  V_{\ms}^{1'2'}
  =4\mathcal{P}\frac{\phi^{1'2'3''4''}\otfsp_{4''2''}\phi^{3''2''3'4'}}
	{2\mu-(\Eno_{3''}+\Eno_{4''})}\mc_{3'4'}, \label{Vmsren}
\end{equation}
where we dropped the recursive~$V_\ms$ term in the anomalous
coupling~$\Sig{A}$ in order to keep Eq.~\eqref{Vmsren} at second
order. We will discuss the recursive term in Sec.~\ref{ladderT}.

We then recognize that we can write the off-diagonal element of the
HFB propagator~$\Sig A$ as the real part of a many-body $T$ matrix
\begin{eqnarray}
  \Sig A&=&V_{\mc}+V_{\ms}=T_{\mc}(2\mu),	\label{renSigA}
\end{eqnarray}
which is defined by
\begin{equation}
  T^{1'2'3'4'}(\epsilon)=2\phi^{1'2'3'4'}+4\mathcal{P}\frac
	{\phi^{1'2'3''4''}\otfsp_{4''2''}\phi^{3''2''3'4'}}
	{\epsilon-(\Eno_{3''}+\Eno_{4''})}. \label{defTV}
\end{equation}
The energies~$\Eno$ are dressed by the normal and mean-field shifts,
but are not the full quasiparticle energies, because they do not
include the effect of the pairing field.  Contractions of this $T$
matrix with anomalous averages are defined by
\begin{equation}
  T^{1'2'}_m(\epsilon)=T^{1'2'3'4'}(\epsilon)\m_{3'4'}.
\end{equation}

The $T$ matrix defined in Eq.\ \eqref{defTV} is a function of energy
through its last two indices in the sense that its
argument~$\epsilon=\Eno_{3'}+\Eno_{4'}$.
\subsection{Diagonal Potentials}
In this Section, we want to redefine the diagonal potentials
$U_{\fc}$\ant$U_\fs$ as the real parts of $T$ matrices by using the
second-order energy shifts~$\mathcal{P}\{\Gamma\}$. With
$\mathcal{P}\{\Gamma\}$ we here denote the principal-value part of the
collisional terms in Eqs.\ \eqref{defgam}\ant\eqref{defup} according
to Eq.~\eqref{principalspe}. We will begin by considering the
condensate potential.
\begin{eqnarray}
  &&-i\;2U_{\fc}+\mathcal{P}\big\{2\Gamma_{\fc\fs\ofs}+\Gamma_{\fs\fs\fc}\nonumber\\
  &&\qquad\qquad\qquad-2\Gamma_{\fc\ofs\fs}-\Gamma_{\ofs\ofs\fc}\big\}\nonumber\\
  &&=-i\;2U_{\fc}-\mathcal{P}\{\Gamma_{1(1+2\fs)\fc}\} 	\label{Uso}
\end{eqnarray}
We here assume real eigenfunctions for the single-particle energy
basis and do not include any $\Gamma$ terms involving the anomalous
average~$\ms$, because they are at least of order
$\V{bin}^3/(\Delta\epsilon)^2$ according to Eq.~\eqref{mphi}.
The second-order terms that contain only normal fluctuations~$\fs$
are used in Eq.~\eqref{Tmpfs} to rewrite the fluctuation potential~$U_\fs$.
The term in Eq.~\eqref{Uso} can again be written in terms of a
many-body $T$ matrix
\begin{equation}
  U_{\fc}+\frac1{2i}\mathcal{P}\{\Gamma_{1(1+2\fs)\fc}\}=T_{\fc},
\end{equation}
which is given by
\begin{equation}
  T^{1'2'3'4'}(\epsilon)=2\phi^{1'2'3'4'}+4\mathcal{P}
	\frac{\phi^{1'2'3''4''}\otfsp_{4''2''}\phi^{3''2''3'4'}}
	{\epsilon-(\Eno_{3''}+\Eno_{2''})}. \label{defTU}
\end{equation}
The slight difference compared to Eq.~\eqref{defTV} is resolved when
we assume diagonal quasiparticle
populations $P_{\bar1\bar2} = P_{\bar1}\delta_{\bar1\bar1}$ as will be
justified in Sec.~\ref{qke}:
\begin{equation}
  \fsp_{1'2'}=U_{1'}^{\bar1}P_{\bar1}U_{2'}^{\bar1\ast}=\fsp_{2'1'},
\end{equation}
where $U$ is the transformation matrix to the quasiparticle basis.
Alternatively, we note that the $T$ matrix is essentially constant for
energy differences up to the duration of a collision.
Contractions of the $T$ matrix with normal averages are performed
according to
\begin{equation}
  T_f^{1'4'}\equiv T^{1'2'3'4'}
	(\Eno_{3'}+\Eno_{4'})f_{3'2'}.
\end{equation}

We now consider the fluctuation potential~$U_\fs$, again include
the truly second-order energy shifts, and obtain
\begin{eqnarray}
  &&-i\;2U_\fs+\mathcal{P}\{\Gamma_{\fs\fs\ofs}-\Gamma_{\ofs\ofs\fs}\}\nonumber\\
  &&=-i\;2U_\fs-\mathcal{P}\{\Gamma_{1\ofs\fs}\}=T'_\fs,	\label{Tmpfs}
\end{eqnarray}
where we get a different $T$ matrix defined by
\begin{equation}
  T'^{1'2'3'4'}(\epsilon)=2\phi^{1'2'3'4'}+4\mathcal{P}
	\frac{\phi^{1'2'3''4''}\ofsp_{4''2''}\phi^{3''2''3'4'}}
	{\epsilon-(\Eno_{3''}+\Eno_{2''})},	\label{defTmp}
\end{equation}
which does not have a factor of $2$ in the intermediate-population
term~$\ofs$. This difference is due to the fact that the mean
field~$\alpha$ is not bosonically enhanced. If we assume diagonal
population~$\fs_{1'2'}=\delta_{1'2'}\fs_{1'1'}$, which is not a good
approximation in this basis, we reproduce the results
of Ref.~\cite{Proukakis1998b} for the GP equation to second order in the
interaction potential. In particular, the factor of $2$ in their
many-body $T$ matrix in the term corresponding to Eq.~\eqref{Tmpfs}
gets canceled with a negative term from adiabatically eliminating
their triple average.
\subsection{Renormalized Propagators}
Using the $T$ matrices defined in the previous sections, we can now
rewrite the generalized GP propagator~$\Pi$ given in
Eq.~\eqref{defpi} and the generalized Boltzmann propagator~$\Sigma$
given in Eq.~\eqref{defsig}.

The Hamiltonian of the GP equation is now
\begin{equation}
   \Pim N'=\Hzero+\T_{\fc}+2T'_{\fs}-\mu,	\label{defpinr}
\end{equation}
and the anomalous coupling~$\Pim A'$ vanishes, because of the identity
\begin{equation}
  V_{\ms}\alpha^\ast=\left\{T_{\mc}(2\mu)-V_{\mc}\right\}\alpha^\ast=
	\left\{T_{\fc}(2\mu)-U_{\fc}\right\}\alpha.
\end{equation}
This means that the coupling between $\alpha$\ant$\alpha^\ast$
vanishes and they are no longer independent quantities, and that we
can without loss of generality treat~$\alpha$ as real.

When we write the Boltzmann propagator in terms of $T$~matrices, we
can explicitly show that the energy spectrum is gapless; this theory
thus fulfills the Hugenholtz-Pines theorem \cite{Hugenholtz1959a}. The
diagonal part of the propagator~$\Sigma$ is now
\begin{equation}
   \Sig{N}'=\Hzero+2\T_{f^c}+2T'_{\fs}-\mu,	\label{defsignr}
\end{equation}
and the anomalous coupling is
\begin{equation}
  \Sig{A}'=T_{\mc}(2\mu).	\label{defsigar}
\end{equation}

We now show that this renormalized Boltzmann propagator~$\Sigma'$ has a
zero-energy eigenvector, i.\thinspace e., its spectrum is gapless. We
begin by writing the generalized GP equation for the condensate ground
state 
\newcounter{mueqcount}\setcounter{mueqcount}{\value{equation}}
\begin{equation}
  \left\{\Hzero+\T_{\fc}(2\mu)+2T'_{\fs}(2\mu)\right\}\alpha
	=\mu\alpha.\label{mueq}
\end{equation}
This equation can be written in terms of the renormalized GP
Hamiltonian~\eqref{defpinr} as $\Pi_{\cal N}\alpha=0$, where the
energies are now measured relative to the adiabatic chemical
potential~$\mu$. An immediate consequence of Eq.~\eqref{mueq} is that
the quasiparticle ground state
\begin{equation}
  P_\alpha=C\begin{pmatrix}\alpha\\-\alpha^\ast\end{pmatrix}\label{defpal},
\end{equation}
with a normalization constant~$C$, is a zero-energy eigenvector of the
renormalized~$\Sigma'$, because of the
identity
\begin{equation}
  T_{\mc}(2\mu)\alpha^\ast=T_{\fc}(2\mu)\alpha.	\label{Tcancel}
\end{equation}

These zero-energy eigenvectors of the HFB propagator are proportional
to the condensate mean field~$\alpha$. All non-zero-energy
eigenvectors $W_{E_{\bar1}\neq0}$ of $\Sigma$ are thus automatically
orthogonal to the condensate, and we can use the complete set
$\{\alpha, W_{E_{\bar1}\neq0}\}$ as a basis to describe the condensate
interacting with thermal excitations \cite{Fetter1972a}. Other
approaches to finite-temperature theories
\cite{Morgan2000a,Castin1997a} have to explicitly orthogonalize their
bases using projection operators. We discuss this new basis in
Sec.~\ref{qke}.

\subsection{Ladder approximation}
\label{ladderT}
We can extend the second-order $T$ matrices introduced in the previous
sections to include ladder diagrams to all orders by using consistency
arguments. We first note that by keeping the recursive~$V_\ms$ term on
the right side of Eq.~\eqref{Vmsren}, we can extend our definition of
the off-diagonal $T$ matrix to
\begin{equation}
  T^{1'2'3'4'}(\epsilon)=2\phi^{1'2'3'4'}+4\mathcal{P}\frac
	{\phi^{1'2'3''4''}\otfsp_{4''2''}T^{3''2''3'4'}(\epsilon)}
	{\epsilon-(\Eno_{3''}+\Eno_{4''})}. \label{defTl}
\end{equation}
This is the real part of the many-body $T$ matrix in the ladder
approximation.

In the previous section, we showed that the HFB propagator~$\Sigma$ is
gapless with the $T$ matrices defined to second order. If we used the
ladder $T$ in Eq.~\eqref{defTl} on the off-diagonal while keeping the
second-order~$T$ in Eq.~\eqref{defTU} in the diagonal, we would find
an energy gap of third order, because the cancellation in
Eq.~\eqref{Tcancel} only works if the two $T$s are identical. However,
the Hugenholtz-Pines theorem tells us that the full theory should
again be gapless. We thus conclude that we have to upgrade the $T$
matrix on the diagonal of~$\Sigma$ to the ladder approximation as well.

We would like to finish our discussion of the scattering matrices in
terms of the single-particle energy basis with two remarks. First, as
the Liouville-space formulation~\cite{Mukamel} of density-matrix
evolution shows, the scattering should really be formulated in terms
of Liouville-space scattering $\mathcal{T}$ matrices~\cite{Fano1963a},
which can be expressed in terms of Hilbert-space $T$ matrices as
\begin{equation}
  \mathcal{T}=T\otimes\openone+\openone\otimes T^\dag+T\otimes T^\dag
	\label{LiouvilleT}.
\end{equation}
Since we only consider Hilbert-space $T$ matrices, we thus would miss
higher-order terms of the type $T\otimes T^\dag$, even if we included
the full many-body Hilbert-space scattering matrix. The imaginary part
of~$\mathcal{T}$ gives rise to the inelastic rates in the kinetic
equations in Sec.~\ref{qke}.

Second, since the asymptotic states in this scattering problem are
typically trapped harmonic-oscillator states for dilute, trapped
atomic gases, we are strictly speaking dealing with bound-state $R$
matrices \cite{Lane1958a} instead of $T$ matrices.
 \section{Quasiparticle Kinetic Equations}\label{qke}
\subsection{Quasiparticle Basis}\label{qbasis}

We now want to write the kinetic equations \eqref{gbe} in the
Bogoliubov quasiparticle basis. In this basis, the complicated and
non-linear evolution due to the HFB propagator~$\Sigma'$ is replaced
with a simple commutator with the eigenenergies and a slow basis
rotation. As this theory is gapless, the $E\neq0$ quasiparticle states
together with the condensate~$\alpha$ form an orthogonal basis,
i.\thinspace e., the thermal fluctuations are by definition orthogonal
to the condensate. Another motivation for transforming to a diagonal
first-order Hamiltonian can be found in the numerical results of
Walser\ea; they find that the reversible first-order evolution leaves
the quasiparticle populations constant. Thus, in the quasiparticle
basis, only the second-order collisional terms change the populations.
A more detailed account of the transformation to the quasiparticle
basis can be found in Ref.~\cite{Wachter2000a}.

Since the quasiparticles consist of the eigenvectors of the
self-energy propagator~$\Sigma'$, we consider the propagator's
$2n$~by~$2n$ ($n$ is the number of single-particle states considered)
eigenvector matrix~$W$ defined by
\newcounter{defWcount}\setcounter{defWcount}{\value{equation}}
\begin{equation} 
  \Sigma' W=WE \label{defW}
\end{equation}
at each time with the diagonal quasiparticle eigenenergy matrix~$E$,
which is labeled with the quasiparticle indices~$\bar1\equiv E_{\bar1}$.
The eigenvalue equations~\eqref{defW} are the Bogoliubov-de-Gennes
equations.  We decompose their solution~$W$ into two $n$~by~$2n$
matrices $U$~and~$V^\ast$,
\begin{equation}
  W=\left(\begin{array}{c}U\\V^\ast\end{array}\right)
	=\begin{pmatrix}u_+&u_-\\v_+^\ast&v_-^\ast\end{pmatrix},
\end{equation}
which are in turn split into $n$~by~$n$ matrices for positive ($u_+$,
$v_+^\ast$) and negative quasiparticle energies ($u_-$, $v_-^\ast$).
These quasiparticle eigenenergies~$\bar1$ are the column indices and
the original single-particle energies~$1'$ the row indices, such that
$U=\left[U^{\bar1}_{1'}\right]_{\{1',\bar1\}}$.

Since the propagator~$\sigma_3\Sigma'$ is positive semi-definite,
the eigenvalues~$E$ in Eq.~\eqref{defW} are real and come in positive
and negative pairs of equal magnitude \cite{Blaizot}. We argued in the
previous section that there exists a zero-energy eigen
vector~$P_\alpha$ defined in Eq.~\eqref{defpal}. This means that the
propagator~$\Sigma'$ is degenerate and has a two-dimensional null space.
The eigenvalue equation~\eqref{defW}, however, does not yield the
second linearly independent zero-energy eigenvector. Instead, the
associated vector~$Q_\alpha$ is given by \cite{Blaizot}
\begin{equation}
  \Sigma' Q_\alpha=-i\frac{P_\alpha}{M},	\label{defq}
\end{equation}
with a positive constant~$M$. In our case, we find
\begin{equation}
  Q_\alpha=-iC\begin{pmatrix}\alpha\\\alpha^\ast\end{pmatrix},
	\label{qsol}
\end{equation}
up to a normalization constant~$C$. In order to find a complete set
of basis states, we now define the quadrature components of
$P_\alpha$\ant$Q_\alpha$:
\begin{subequations}
\begin{eqnarray}
  W^{+0}&\equiv&\frac1{\sqrt{2}}(P_\alpha+iQ_\alpha)=
	\sqrt{2}C\begin{pmatrix}\alpha\\0\end{pmatrix},\\
  W^{-0}&\equiv&-\frac1{\sqrt{2}}(P_\alpha-iQ_\alpha)=
	\sqrt{2}C\begin{pmatrix}0\\\alpha^\ast\end{pmatrix}.
\end{eqnarray}
\end{subequations}
If we now substitute these two states $W^{+0}$\ant$W^{-0}$ into the
zero-energy columns of~$W$, we can normalize all eigenvectors using
the symplectic norm
\begin{equation}
  W^\dag\sigma_3W=\sigma_3.	\label{Wnorm}
\end{equation}
The negative-energy states thus have negative norm, and the
positive-energy states positive norm. In particular, the zero-energy
vector~$W^{+0}$ has positive norm and thus belongs to the
positive-energy part of~$W$.  This normalization fixes the
constant~$C$ for the zero-energy states to
\begin{equation}
  C=\frac1{\sqrt{2\alpha^\dag\alpha}}\equiv\frac1{\sqrt{2N_c}},
\end{equation}
where we defined the number of condensate atoms~$N_c$.  The
completeness relation for the quasiparticle basis also has symplectic
structure
\begin{equation}
  W\sigma_3W^\dag\sigma_3=\openone_{2n},	\label{Wcompl}
\end{equation}
where~$\openone_{2n}$ indicates the $2n$-dimensional unit-matrix.
The symplectic structure of the orthonormalization and completeness
relations Eqs.\ \eqref{Wnorm}\ant\eqref{Wcompl}, \ie the appearance of
the matrix~$\sigma_3$ in these expressions, guarantees that the
transformation to the quasiparticle basis is canonical \cite{Blaizot}.
Canonical here means that the new quasiparticle operators obey the
boson commutation relations.

To further examine the structure of the quasiparticle states~$W$, we
consider the following symmetry of the HFB propagator~$\Sigma'$
\begin{equation}
  \Sigma'=-\sigma_1\Sigma'^\ast\sigma_1,\\
\end{equation}
which holds according to its definition in Eq.~\eqref{defsig}.
This symmetry implies \cite[IV.B.]{Walser2001a} the following
relation for the quasiparticle states~$W$:
\begin{equation}
  W=\begin{pmatrix}u_+&u_-\\v_+^\ast&v_-^\ast\end{pmatrix}
	=\begin{pmatrix}u_+&v_+\\v_+^\ast&u_+^\ast\end{pmatrix}. 
	\label{Wminus}
\end{equation}

This relation in Eq.~\eqref{Wminus} shows that the positive- and
negative-energy eigenvectors of~$\Sigma'$ are not independent but
related by
\begin{equation}
  W^{\bar1}=\sigma_1W^{-\bar1\ast}.	\label{Wpm}
\end{equation}
Plugging $W$ from Eq.~\eqref{Wminus} into the full completeness
relation Eq.~\eqref{Wcompl}, we find a completeness relation for the
independent elements of~$W$ alone:
\begin{equation}
  \frac1{N_c}\alpha\alpha^\dag+\sum_{\bar1>0}(u_+^{\bar1}u_+^{\bar1\ast}-
	v_+^{\bar1}v_+^{\bar1\ast})=\openone_n.	\label{completeuv}
\end{equation}
This is the basis we want to use for the kinetic equations. The
explicit split in a condensate mode~$\alpha$ and the orthogonal
fluctuation modes $u^{\bar1>0}$\ant$v^{\bar1>0}$ is similar to the
formalism of Ref.~\cite{Fetter1972a}. While the condensate
mode~$\alpha$ evolves according to Eq.~\eqref{ggpe} with the updated
propagator~$\Pi'$, we have to find an evolution equation for the
quasiparticle populations.

We thus write the generalized single-time fluctuation-density
matrix~$\Gs^<$ defined in Eq.~\eqref{def_gsl} in terms of
quasiparticles as
\begin{equation}
  \Gs^<=WPW^\dag=\left(\begin{array}{cc}
		UPU^\dag& UPV^\top\\V^\ast PU^\dag& V^\ast PV^\top
		\end{array}\right),	\label{defGg}
\end{equation}
where~$P$ is the not necessarily diagonal $2n$~by~$2n$ quasiparticle
population matrix.  The time-reversed density matrix~$\Gs^>$
transforms according to its definition Eq.~\eqref{def_gsg}.  From the
same equation, we can also deduce that the quasiparticle population
matrix~$P$ is Hermitian and fulfills a similar identity:
\begin{equation}
  \sigma_3+P=\sigma_1P^\ast\sigma_1=\sigma_1P^\top\sigma_1.	\label{Pid}
\end{equation}
This identity implies that~$P$ can be written as
\begin{equation}
  P=\begin{pmatrix}p\;&q\\q^\ast&(1+p)^\ast\end{pmatrix},
	\label{Ppm}
\end{equation}
in a structure similar to~$\Gs^<$ in Eq.~\eqref{def_gsl}. Since the
classical mean field~$\alpha$ only undergoes stimulated emission, the
ground-state factors~$P_{+0+0}=P_{-0-0}$ do not contain enhancement
terms. Thus, the zero-energy components of $\sigma_3$ in
Eq.~\eqref{Pid} and similar enhancement terms associated
with~$\alpha$ actually have to be zero.

We also represent the condensate~\eqref{defchi} by a population
matrix~$P^c$ in the quasiparticle basis
\begin{equation}
  \chi\chi^\dag=WP^cW^\dag.	\label{defPc}
\end{equation}

In order to maintain orthogonality between the condensate and the
thermal excitations, we have to demand an analogue to the
constraint~\eqref{alphaGort} in the quasiparticle basis. We here have
to distinguish two cases.

In the first case, the system evolves slowly enough that the
ground-state of the adiabatic basis given in Eq.~\eqref{completeuv} is
the true condensate state~$\alpha(t)=\alpha$ (real). We can then
explicitly give the condensate matrix as
\begin{equation}
  P^c=N_c e^{\vphantom{\dag}}_{0}e_{0}^{\dag},	\label{adPc}
\end{equation}
with a vector~$e_{0}^{\bar1}=\delta_{\bar1,\pm0}$, such that all
non-zero-energy components of $P_c$ vanish.  To implement the
orthogonality constraint, we then explicitly set the $E=\pm0$ elements of
the quasiparticle matrix $P$ to zero:
\begin{equation}
  P_{0\bar1}=P_{\bar10}=0,\quad\text{for all }\bar1.	\label{staticort}
\end{equation}

In the second case, the system evolves too fast for the quasiparticle
basis to follow. Then, the condensate matrix contains components of
non-zero energy. In this case, the more general orthogonality
constraint
\begin{equation}
  \bra{\alpha(t)}P=\scp{\alpha(t)}{\bar1}P_{\bar1\bar2}\bra{\bar2}=0
	\label{dynort}
\end{equation}
has to be fulfilled. This constraint is particularly important when
the condensate is coherently excited in linear response. For adiabatic
evolution, Eq.~\eqref{dynort} reduces to Eq.~\eqref{staticort}.  In
the next section, we will find the evolution equation for $P$.

\subsection{Kinetic Equations}
With these ingredients, we can now try to obtain an equation of motion
for the occupation number matrix~$P$ from the kinetic
equation~\eqref{gbe}. We use Eq.~\eqref{defGg} to substitute the
fluctuation-density matrices~$\Gs$ and obtain
\begin{eqnarray}
  \der{t}P&=&-i[E,P]_-+\Big\{W^{-1}\frac{dW}{dt}P+\hc\Big\}\\
	&+&\Big\{W^{-1}\Gamma^<\;W(\sigma_3+P)
 	-W^{-1}\Gamma^>WP+\hc\Big\}. \nonumber\label{Peq}
\end{eqnarray}
We are now left with the task of transforming the collisional
contributions to the quasiparticle basis. To this end, we define new
two-particle matrix elements in the quasiparticle energy basis by
\begin{equation}
  \Phi^{\bar1\bar2\bar3\bar4}\equiv
	\phi^{1'2'3'4'}\;V^{\bar1\ast}_{1'}V_{2'}^{\bar2\ast}
	U_{3'}^{\bar3}U_{4'}^{\bar4}.
	\label{defPhi}
\end{equation}
These quasiparticle matrix elements have the same symmetries as the
original ones given in Eq.~\eqref{phisymm}. Furthermore, they fulfill
\begin{equation}
  \Phi^{\ast\bar1\bar2\bar3\bar4}\equiv
  \big(\Phi^{\bar1\bar2\bar3\bar4}\big)^\ast
    	=\Phi^{-\bar3-\bar4-\bar1-\bar2}.\label{Phiid}
\end{equation}

The careful examination of symmetries in Ref.~\cite{Wachter2000a} shows
that the collision operator~Eq.~\eqref{defgam} reduces to terms
containing the completely symmetric matrix elements
\begin{eqnarray}
  \psi^{\bar1\bar2\bar3\bar4}\equiv	\label{defpsi}
	\frac1{\sqrt3}\sum_\pi\;\Phi^{\pi(\bar1\bar2\bar3\bar4)},
\end{eqnarray}
which are defined as a sum over all index permutations~$\pi$.

The resulting kinetic equations for the quasiparticle populations~$P$ are
\newcounter{mastercount}\setcounter{mastercount}{\value{equation}}
\begin{eqnarray} 
  \der{t}P=-iEP+B_WP+C_{PP}+C_{\alpha P}+\hc,
	\label{master}
\end{eqnarray}
with the non-adiabatic basis-rotation term
\begin{equation}
  B_W=W^{-1}\frac{dW}{dt}=\sigma_3W^\dag\sigma_3\frac{dW}{dt},
  \label{defBW}
\end{equation}
and the Boltzmann collision rates between fluctuations
\begin{eqnarray}
  C_{PP}&=&C_{PP}^<(\sigma_3+P)-C_{PP}^>P\label{defCPP}\\
	&=&\Gamma_{PP\opp}\opp-\Gamma_{\opp\opp P}P\nonumber
\end{eqnarray}
and between fluctuations and the condensate
\begin{equation}
  C_{\alpha P}=3\Gamma_{P^cP\opp}\opp-3\Gamma_{P^c\opp P}P. \label{defCaP}
\end{equation}
These collision rates are defined in terms of the quasiparticle
collision operator
\begin{equation}
  \sigma_3\Gamma_{PPP}=\frac12
	\psi^{\bar1\bar2\bar3\bar4}\psi_\eta^{\bar1'\bar2'\bar3'\bar4'}
	\Pv_{-\bar1-\bar1'}\Pv_{-\bar2-\bar2'}\Pv_{\bar4'\bar4}
	\ket{\bar3}\otimes\bra{\bar3'}.
\end{equation}
In this operator, $P$ can stand for any one of the three possibilities
$P$, $\opp$, or $P^c$, as needed in Eqs.\
\eqref{defCPP}\ant\eqref{defCaP}.
The approximately energy conserving matrix element~$\psi_\eta$ is
explicitly given by
\begin{equation}
  \psi_\eta^{\bar1'\bar2'\bar3'\bar4'}=\psi^{\bar1'\bar2'\bar3'\bar4'}
	\pi\delta_\eta(E_{\bar1'}+E_{\bar2'}+E_{\bar3'}+E_{\bar4'}),
	\label{psieta}
\end{equation}
where the quasiparticle energies can be positive or negative depending
on their index. In an on-shell scattering event, this delta function
forces two of the indices to be positive and the remaining two to be
negative. This has to be considered in interpreting the collision
terms $C_{PP}$\ant$C_{\alpha P}$, because there all the sums run over
positive and negative indices. The principal-value part in
Eq.~\eqref{principalspe}, which appears in the single-particle kinetic
equations, is absorbed in the $T$ matrices in Sec.~\ref{ren} and thus
does not appear anymore in the quasiparticle equations.
\begin{figure}[h]
  \centering\includegraphics[scale=.8]{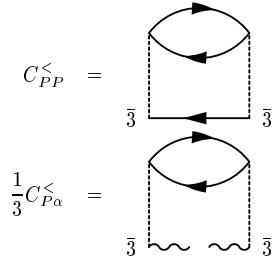}
  \caption{The incoming collision rates for collisions between thermal
  atoms~$C_{PP}^<$ and between a thermal and a condensate
  atom~$C_{\alpha P}^<$. The tight potential lines are now totally
  symmetric according to Eq.~\eqref{defpsi}. The propagator lines
  represent the quasiparticle propagator~$P$, which contains both the
  anomalous average~$\ms$ and the normal density~$\fs$. Because of the
  total symmetry of the interaction line, the three distinct
  condensate diagrams on each row of Fig.~\ref{fig_gammasq} reduce to
  one diagram with a weight of 3. }
  \label{fig_gammaP}
\end{figure}

To complete the presentation of the quasiparticle kinetic equations,
we write the generalized Gross-Pitaevskii equation~\eqref{ggpe} with
the updated GP propagator~$\Pi'$ and the
second-order collisional contributions expressed in terms of
quasiparticle populations~$P$ and obtain
\newcounter{Pceqcount}\setcounter{Pceqcount}{\value{equation}}
\begin{equation} 
  \der{t}\chi=-i\Pi'\chi+W\sigma_3
	\big(\Gamma_{PP(\sigma_3+P)}-\Gamma_{(\sigma_3+P)(\sigma_3+P)P}\big)
	W^{-1}\chi
  \label{Pceq}
\end{equation}
In Sec.~\ref{summ}, we write this equation for $\alpha(t)$ alone.
This equation together with Eq.~\eqref{master} gives a complete
description in terms of Bogoliubov quasiparticles of a condensate
coupled to a thermal cloud at finite temperatures. 
\subsection{Orders of magnitude}
We now want to discuss the orders of magnitude of several of the
quantities of this theory. This will suggest some approximations to the
full quasiparticle kinetic equations~\eqref{master}.

We first consider the basis transformation $W$. The completeness
relation for the quasiparticle basis Eq.~\eqref{Wcompl} tells us that
\begin{equation}
  W=\LandO(1)	\label{Wmag}.
\end{equation}
For example, for high-energy eigenfunctions, the effect of the
condensate becomes small, and the quasiparticle transformation reduces
to
\begin{equation}
  u\longrightarrow1\qandq v\longrightarrow0,\quad
	\text{as}\;E\longrightarrow\infty.	\label{uvmag}
\end{equation}

Now, considering the basis-rotation term defined in Eq.~\eqref{defBW}
we find
\begin{equation}
  B_W=W^{-1}\frac{dW}{dt}=\frac{\LandO(1)}{dt}=\LandO(\Gamma)<
	\LandO\Big(\frac1{\tau_c}\Big),	\label{BWmag}
\end{equation}
because the time-scale for population changes, which change the
quasiparticle basis~$W$, is limited by the time between
collisions~$dt>\tau_c$. In equilibrium, the populations are constant
due to detailed balance of the in and out rates. Thus, the net
collision rate~$\Gamma\equiv C_{PP}+C_{\alpha P}+\hc$ gives a better estimate
for population changes~$dt\approx\Gamma^{-1}$. This also confirms
that $B_W=0$ in equilibrium, since $dW/dt=0$.

We now show that the stationary solutions~$P$ of the Boltzmann
equation~\eqref{master} are diagonal.  Considering the stationary
solution~$\der{t}P_{\bar1\bar2}=0$ of Eq.~\eqref{master} for an
off-diagonal element with~$\bar1\neq\bar2$, we obtain
\begin{equation}
  P_{\bar1\neq\bar2}=i\frac{B_WP+C_{PP}+C_{\alpha P}+\hc}{E_{\bar2}-E_{\bar1}}
	=\LandO\Big(\frac{\Gamma}{\De}\Big).\label{Podsmall}
\end{equation}
This shows that the off-diagonal elements of the quasiparticle
population are small compared to the diagonal ones, which are of the
order of the number of particles~$N$, and vanish at equilibrium.

\subsection{Quasiparticle $T$ matrices}

The $T$ matrix in Eq.~\eqref{defTV} (or the ladder extension
Eq.~\eqref{defTl}) has been obtained appropriately to second
order. The second-order term with the factor of 1 in Eq.~\eqref{defTV}
is the divergent part, which is renormalized in Eq.~\eqref{defTl} when
replaced by the $T$-matrix. The term containing $2\fs$ of
Eq.~\eqref{defTV} is a convergent many-body second-order term. It is
thus reasonable to replace the $T$ of Eq.~\eqref{defTl} with
\begin{equation}
  T^{1'2'3'4'}(E)=T_{2B}^{1'2'3'4'}(E+2\mu)+8\mathcal{P}
	\frac{\phi^{1'2'3''4''}\fs_{4''2''}\phi^{3''2''3'4'}}
	{E-(E_{3''}+E_{4''})},	\label{defTqp}
\end{equation}
where we replaced the single-particle energies in the denominator by
quasiparticle energies, because the difference is of higher order.
We here use the following two-body $T$~matrix
\begin{equation}
  T_{\text{2B}}^{1'2'3'4'}(\epsilon)=2\phi^{1'2'3'4'}+4\mathcal{P}\frac
	{\phi^{1'2'3''4''}\phi^{3''4''3'4'}}
	{\epsilon-(\Esp_{3''}+\Esp_{4''})}, \label{defT2B}
\end{equation}
which is given in terms of single-particle energies~$\Esp$ defined by
\begin{equation}
  \Hzero\ket{\Esp_{1'}}=\Big(\frac{\hat{\bp}^2}{2m}+\V{ext}(\hat{\bx})\Big)
	\ket{\Esp_{1'}}=\Esp_{1'}\ket{\Esp_{1'}}.	\label{defesp}
\end{equation}

The collisional terms of the kinetic equations~\eqref{master}
correspond to the imaginary part of a Liouville-space
$\mathcal{T}$-matrix~\eqref{LiouvilleT}.  Examining the argument of
the $\delta_\eta$-function in Eq.~\eqref{psieta}, which
defines~$\Psi_\eta$, in comparison with the ladder $T$ in
Eq.~\eqref{defTl} shows that (when the energies are correlated with
the appropriate elements of W, see \cite{Wachter2000a}) two of the
energies in the denominator must be positive and two negative, which,
together with the fact that $P$ is diagonal close to equilibrium,
leads to terms in the kinetic equation of the form $PP(1+P)(1+P)$,
etc. The kinetic equation~\eqref{master} then becomes the quantum
Boltzmann equation for quasiparticle populations. The equilibrium
solution is therefore the expected Bose-Einstein distribution for the
quasiparticles, as the steady-state solution of Eqs.\
\eqref{master}\ant\eqref{Pceq} shows \cite[Sec.~V]{Walser2001a}. The
interaction matrix elements~$\phi$ in Eq.~\eqref{defPhi} can also be
upgraded to $T$s. These equations in terms of $T$ are now consistent
with an impact approximation treatment (with elastic scattering not
contributing when $T\otimes T^\dag$ terms are considered, cf
Eq.~\eqref{LiouvilleT}).

\subsection{Conservation laws}
Because of the non-vanishing pairing field, the trace of the
quasiparticle density matrix~$P$ is in general not equal to the number
of excited particles in the system. Hence, the number of
quasi-particles is not conserved.
Our single-particle kinetic equations \eqref{ggpe}\ant\eqref{gbe}
do, however, conserve the mean total number
\begin{equation}
  \ave{N}=\alpha^\dag\alpha+\tr\{\fs\}=N_c+\tr\{\fs\}=\mathrm{const};
\end{equation}
this can be proven explicitly by plugging these kinetic equations
into~$\der{t}\ave{N}$ and canceling terms.  We thus adopt a
self-consistent procedure for the quasiparticle kinetic
equations Eqs.\ \eqref{master}\ant\eqref{Pceq}, by requiring
\begin{equation}
  \ave{N}=N_c+\tr\{UPU^\dag\}=\mathrm{const}.	\label{Nconstqp}
\end{equation}
This equation self-consistently constrains the number of condensate
atoms~$N_c$: in equilibrium at temperature~$\beta^{-1}$, the
quasiparticle matrix $P$ consists of $q=0$ and
\begin{equation}
 p(E)=\frac1{e^{\beta(E-\mu)}-1}
\end{equation}
according to Eq.~\eqref{Ppm} with the chemical potential~$\mu$ given
by the GP equation~\eqref{mueq}. As temperature tends to zero, we
obtain the usual corrections for the anomalous average and condensate
depletion \cite{Morgan2000a}.  Note that the number of excited atoms
is not given by the trace of~$p$, but has to include the basis
transformation~$U$ as discussed above.  If we drop the basis-rotation
term~$B_W$ in numerical simulations, we will incur
number-non-conservation on the order of~$\Gamma$, while away from
equilibrium.

Since we use a Markovian collision integral with a damping function of
finite width~$\eta$ in order to include off-the-energy-shell
propagation, this theory is not exactly energy conserving. Markovian
theories fail to track the decay of initial correlations and thus do
not account for the decay of the correlation energy \cite{Bonitz}.  In
our case, with a self-consistent~$\eta$, energy is conserved to
order~$\eta$, which is consistent with the order of
approximation.  For a detailed discussion of these memory effects and
how they affect the conservation laws see \cite{Bhongale2002a}.

\section{Summary}\label{summ}
We summarize the main results of this paper given in Secs.\ 
\ref{ren}\ant\ref{qke}. We formulate a kinetic theory in terms of
Bogoliubov quasiparticle modes~$W$, which are defined by the
eigenvalue equation for the renormalized Hartree-Fock-Bogoliubov
propagator $\Sigma'$
\newcounter{tmpcount}\setcounter{tmpcount}{\value{equation}}
\setcounter{equation}{\value{defWcount}}
\begin{equation}
  \begin{pmatrix}\Hzero+2\T_{f^c}+2T'_{\fs}-\mu& T_{\mc}\\
	-\Sig A'^\ast& -\Sig N'^{\ast}\end{pmatrix}W=WE.
\end{equation}
The $T$ matrices are defined to second order in
Eqs.\ \eqref{defTV}\ant\eqref{defTmp} and Eq.~\eqref{defTl} gives an
extension to the ladder approximation.
The Gross-Pitaevskii equation
\setcounter{equation}{\value{mueqcount}}
\begin{equation}
  \left\{\Hzero+\T_{\fc}(2\mu)+2T'_{\fs}(2\mu)\right\}\alpha
	=\mu\alpha
\end{equation}
for the ground state~$\alpha$ is contained in Eq.~\eqref{defW},
because the renormalized~$\Sigma'$ is gapless. The GP equation defines
the value of the chemical potential~$\mu$. To find a complete
basis~$W$, we have to find the second zero-energy mode and form
quadrature components as discussed in Sec.~\ref{qbasis}.

We find the following Boltzmann equation for the thermal excitations in
terms of Bogoliubov quasiparticles:
\setcounter{equation}{\value{mastercount}}
\begin{eqnarray}
  \der{t}P&=&-i[E,P]+\Big\{\sigma_3W^{\dag}\sigma_3\frac{dW}{dt}P+\hc\Big\}\\
	&+&\Big\{\Gamma_{PP\opp}\opp-\Gamma_{\opp\opp P}P+\hc\Big\}\nonumber\\
 	&+&3\Big\{\Gamma_{P^cP\opp}\opp-\Gamma_{P^c\opp P}P+\hc\Big\}.\nonumber
\end{eqnarray}
The basis-rotation term containing~$\der{t}W$ can be dropped for
adiabatic evolution. The quasiparticle density matrix~$P$ is diagonal
close to equilibrium, and its elements obey an equilibrium
Bose-Einstein distribution as the second and third lines show. The
collision terms containing the general condensate matrix~$P^c$ defined
in Eq.~\eqref{defPc} represent population exchange between the thermal
cloud and the condensate. They are balanced in the following equation
for the condensate \setcounter{tmpcount}{\value{equation}}
\setcounter{equation}{\value{Pceqcount}}
\begin{eqnarray}
  \der{t}\alpha(t)&=&
	\big\{\Hzero+\T_{\fc}+2T'_{\fs}-\mu\Big\}\alpha(t)\\
	&+&U\sigma_3\big\{\Gamma_{PP(\sigma_3+P)}-
	\Gamma_{(\sigma_3+P)(\sigma_3+P)P}\big\}\sigma_3\nonumber\\
	&&\times\big\{U^\dag\alpha(t)-V^\top\alpha^\ast(t)\big\}.\nonumber
\end{eqnarray}
\setcounter{equation}{\value{tmpcount}}In general, $\alpha(t)$ can be
different from the $\alpha$ used as the ground state of the adiabatic
basis. A non-adiabatic change in a driving force, for example, would
cause $\alpha(t)$ to change quickly.  

The coupled Eqs.\ \eqref{master}\ant\eqref{Pceq} have to be solved
under the orthogonality constraints
\eqref{dynort}~or~\eqref{staticort} depending on whether the evolution
is adiabatic or not. Furthermore, the total particle number has to
fulfill the self-consistent constraint Eq.~\eqref{Nconstqp}.
\section{Conclusions}
We have extended the non-equilibrium kinetic theory of Walser\ea\ 
\cite{Walser1999a, Walser2001a} in two important
respects. First, we write the binary interactions as many-body $T$
matrices to second order in the interaction by subsuming ultra-violet
divergent terms. This procedure removes divergences caused by the
anomalous average~$\ms$ and contained in the second-order terms. We
can then replace the low-energy limit of the $T$ matrix with the
$s$-wave scattering length. This is to our knowledge the only
consistent treatment of these difficulties associated with the
anomalous average in a theory that includes second-order collisional
terms. The updated Hartree-Fock-Bogoliubov propagator~$\Sigma'$ is
then gapless, which greatly facilitates a consistent treatment of the
condensate coupled to thermal fluctuations.

The second extension of the Walser\ea\ theory makes use of the
gapless HFB propagator by using its quasiparticle eigenmodes to
parameterize the thermal fluctuations, which are then automatically
orthogonal to the condensate mean field. We find that this basis
greatly simplifies the second-order collision terms of the Walser\ea\ 
theory. Another important result reported here is that, in 
equilibrium, the Bogoliubov modes diagonalize the quasiparticle
population matrix~$P$. This means that, close to equilibrium, the
second-order terms can be evaluated in $n^4$ operations, where $n$ is
the number of energy levels considered. This is a vast improvement
compared to $n^8$ operations for a basis that does not diagonalize the
population matrix, such as, for example, the single-particle basis
used by Walser\ea

\section*{Acknowledgments}
J.W.\ acknowledges financial support from the National Science
Foundation.  R.W.\ gratefully acknowledges support from the Austrian
Academy of Sciences through an APART grant. M.H.\ acknowledges support
from the U.S.\ Department of Energy, Office of Basic Energy Sciences
via the Chemical Sciences, Geosciences, and Biosciences
Division. J.C.\ is retired and acknowledges support from TIAA-CREF.


\end{document}